\documentstyle{mn}

\newif\ifAMStwofonts

\input epsf

\def\lesssim{\mathrel{\hbox{\rlap{\hbox{\lower4pt\hbox{$\sim$}}}\hbox{$<$}}}}

\def\gtrsim{\mathrel{\hbox{\rlap{\hbox{\lower4pt\hbox{$\sim$}}}\hbox{$>$}}}}

\def\msun{M$_{\odot}$}

\def\teff{$T_{\rm eff}$~}

\def\ll_lsun{$\log({L/\rm L_{\odot}})$~}

\def\masa_msun{$M/ \rm M_{\odot}$~}

\def\m_mstar{$M/M_{*}$~}

\title{The ages and colours of cool helium-core white dwarf stars}

\author [A. M. Serenelli, L. G. Althaus, R. D. Rohrmann and 
O. G. Benvenuto]
{A. M. Serenelli$^1$\thanks{Fellow of the
Consejo Nacional de
Investigaciones Cient\'{\i}ficas y T\'ecnicas (CONICET), Argentina.
Email: serenell@fcaglp.fcaglp.unlp.edu.ar}
L.   G. Althaus$^1$\thanks{Member of the Carrera del Investigador
Cient\'{\i}fico y Tecnol\'ogico, Consejo Nacional de
Investigaciones Cient\'{\i}ficas y T\'ecnicas (CONICET), Argentina.
Email: althaus@fcaglp.fcaglp.unlp.edu.ar} 
R. D. Rohrmann$^2$\thanks{Fellow of the
Consejo Nacional de
Investigaciones Cient\'{\i}ficas y T\'ecnicas (CONICET), Argentina.
Email: rohr@oac.uncor.edu} and
O. G. Benvenuto$^1$\thanks{Member  of  the  Carrera  del Investigador
Cient\'{\i}fico, Comisi\'on  de Investigaciones  Cient\'{\i}ficas
de la Provincia de Buenos Aires, Argentina. Email:
obenvenuto@fcaglp.fcaglp.unlp.edu.ar} \\
$^1$Facultad  de  Ciencias
Astron\'omicas y Geof\'{\i}sicas, Universidad Nacional de La
Plata, Paseo del Bosque S/N, (1900) La Plata, Argentina \\
$^2$Observatorio Astron\'omico, Universidad Nacional de
C\'ordoba, Laprida 854, (5000) C\'ordoba, Argentina}

\date{November 13}

\pagerange{\pageref{firstpage}--\pageref{lastpage}}

\pubyear{2000}

\begin{document}

\maketitle

\label{firstpage}

\begin{abstract}  

The purpose  of this work is  to explore the  evolution of helium-core
white dwarf  stars in  a self-consistent way  with the  predictions of
detailed non-gray  model atmospheres  and element diffusion.   To this
end, we consider helium-core white dwarf models with stellar masses of
0.406, 0.360,  0.327, 0.292, 0.242,  0.196 and 0.169 \msun  and follow
their  evolution from  the  end  of mass  loss  episodes during  their
pre-white dwarf evolution down to very low surface luminosities.

We find that when the effective temperature decreases below 4000K, the
emergent spectrum  of these stars becomes bluer  within time-scales of
astrophysical interest. In particular, we analyse the evolution of our
models in the colour-colour  and colour-magnitude diagrams and we find
that helium-core white dwarfs with  masses ranging from $\sim$ 0.18 to
0.3 M$_{\odot}$  can reach  the turn-off in  their colours  and become
blue again within cooling times much  less than 15 Gyr and then remain
brighter  than $M_V  \approx$ 16.5.   In view  of these  results, many
low-mass helium white  dwarfs could have had time  enough to evolve to
the  domain of collision-induced  absorption from  molecular hydrogen,
showing blue colours.

\end{abstract}

\begin{keywords}  stars:  evolution  -  stars: interiors - stars:
white dwarfs - stars: atmospheres - stars: fundamental parameters

\end{keywords}

\section{Introduction} \label{sec:intro}

The theoretical and observational study of white dwarf (WD) stars with
helium  cores  (hereafter  He  WD)  is a  subject  that  has  received
increased  attention particularly during  the last  few years.   It is
well  known for  instance that  close binary  evolution and  mass loss
episodes are  required to form low-mass  He WDs within the  age of the
Galaxy (see Iben  \& Livio 1993 for a review).   These He WDs populate
the tail  of low-mass ($M < $  0.4 \msun) in the  WD mass distribution
(Bergeron, Saffer \& Liebert 1992; Bragaglia, Renzini \& Bergeron 1995
and  Saffer, Livio  \& Yungelson  1998).  The binary  nature of  these
objects was first placed on a firm observational basis by Marsh (1995)
and  Marsh, Dhillon  \& Duck  (1995).  Since  then, He  WDs  have been
detected in numerous binary configurations involving either another WD
or a neutron  star (see, e.g., Lundgren et al.   1996; Moran, Marsh \&
Bragaglia 1997; Orosz et al.   1999; Maxted et al.  2000).  Several He
WDs have  also been found in  open and globular  clusters (Landsman et
al.  1997; Cool et al.   1998 and Edmonds et al.  1999).  Evolutionary
models  appropriate  for  the  study  of He  WDs  have  recently  been
presented by Althaus \& Benvenuto (1997), Benvenuto \& Althaus (1998),
Hansen \& Phinney (1998a) and Driebe et al. (1998).

The  determination of  ages  for  low-mass He  WDs  is a  longstanding
problem in the study of the  evolution of these stars.  As a matter of
fact, cooling ages for He WD are intimately related to the mass of the
hydrogen envelope left before entering the final cooling track.  Early
investigations  carried out  by Webbink  (1975) indicated  that,  as a
result  of massive  hydrogen envelopes,  hydrogen burning  via  the pp
cycle remains  the main  important source of  energy over most  of the
computed evolution, thus  implying very long evolutionary time-scales.
In agreement  with this prediction,  Alberts et al. (1996)  and Sarna,
Antipova  \& Ergma  (1999)  found that,  on  the basis  of a  detailed
treatment of  the binary evolution  leading to the formation  of these
objects,  the  hydrogen-rich  envelope  surviving  flash  episodes  is
massive enough  for residual hydrogen  burning to be dominant  even at
low luminosities, thus resulting  in very long cooling times.  Massive
hydrogen  envelopes for  He WDs  have also  been derived  by  Driebe et
al. (1998),  who have  simulated the binary  evolution by forcing  a 1
\msun  model at  the  red giant  branch  to a  large  mass loss  rate.
Obviously, very large cooling  ages as derived from these evolutionary
models  constrain the  existence of  He  WDs with  very low  effective
temperature (\teff) values.  For example,  the Driebe et al.  model of
0.195 \msun needs  about 17 Gyr to evolve down  to \teff= 5000K, which
exceeds any age range of  interest.  The possibility that the hydrogen
envelope is considerably reduced  by additional mass loss episodes was
considered by  Iben \& Tutukov  (1986).  These authors found  that the
last flash  episode undergone by their  0.3 \msun He  WD model becomes
strong  enough  so  as  to  force  the  model  to  reach  again  giant
dimensions,  causing   Roche-lobe  overflow.   They   found  that  the
resulting  mass transfer  reduces  the hydrogen  envelope  to such  an
extent  that it  is unable  to  support any  further nuclear  burning,
thereby   implying   short   cooling    ages   at   late   stages   of
evolution. Finally, Benvenuto \&  Althaus (1998) and Hansen \& Phinney
(1998a)  presented  a detailed  grid  of He  WD  models  based on  the
assumption  that He  WDs are  formed with  a relatively  thin hydrogen
envelope and thus hydrogen  burning does not appreciably contribute to
the luminosity budget of the star.

In the context of the foregoing discussion, binary systems composed by
a  millisecond   pulsar  and  a   low-mass  He  WD   are  particularly
interesting, as they  may place tight constraints on  the evolution of
He WDs. Indeed, the spin-down age of the pulsar provides an estimation
of the  age of  the He WD  companion, which  can be compared  with the
predictions of evolutionary calculations  for the WD.  In this regard,
the best-studied system belonging to this class is PSR J1012+5307, for
which Lorimer et al. (1995)  determined a spin-down age for the pulsar
($\tau_c=$) of 7  Gyr.  In addition, atmospheric parameters  of the He
WD component have  been derived by van Kerkwijk,  Bergeron \& Kulkarni
(1996) and  Callanan,  Garnavich \&  Koester  (1998) (\teff  $\approx$
8500K and  log g $\approx$  6.3-6.7).  Evolutionary models for  He WDs
that predict  thick hydrogen envelopes  (Driebe et al. 1998)  lead for
this  low-mass WD ($M  \approx$ 0.15-0.19  \msun) to  an age  which is
consistent  with  the spin-down  age  of  the  pulsar. However,  large
evolutionary  time-scales   as  predicted  by  these   models  are  in
disagreement  with   new  observational  data.   In   fact,  a  recent
breakthrough  has  been  the  optical  detection of  the  low-mass  WD
companion to  the millisecond pulsar  PSR B1855+09 by van  Kerkwijk et
al. (2000),  who derived for  the WD companion  a \teff value  of 4800
$\pm$ 800K.   The mass of the  WD in this binary  system is accurately
known   thanks   to  the   Shapiro   delay   of   the  pulsar   signal
(0.258$^{+0.028}_{-0.016}$   \msun;   see   Kaspi,  Taylor   \&   Ryba
1994). According  to the Driebe et  al.' models, this  low \teff value
corresponds to a WD  cooling age of 10 Gyr, which is  at odds with the
characteristic age of the pulsar,  $\tau_c=$ 5 Gyr.  In addition, ages
well above 10 Gyr are inferred from the models of Driebe et al. (1998)
for the very  cool WD companions to PSR  J0034-0534 and PSR J1713+0747
(Hansen  \& Phinney  1998b),  which  may be  an  indication that  such
evolutionary  models  indeed  overestimate  the WD  cooling  ages,  as
discussed by van Kerkwijk et  al. (2000). Finally, a recent population
study of  close double WDs in  the Galaxy carried out  by Yungelson et
al. (2000) suggests that low mass He WDs have to cool much faster than
suggested by evolutinary models of Driebe et al. (1998).

Very  recently,   Althaus,  Serenelli  \&   Benvenuto  (2000a,b)  have
presented new evolutionary calculations  for He WDs aimed at exploring
the  role played  by element  diffusion in  the ocurrence  of hydrogen
shell flashes and, more importantly, to assess whether or not the mass
of  the hydrogen  envelope  can be  considerably  reduced by  enhanced
hydrogen  burning   during  diffusion-induced,  flash   episodes.   In
particular,  He WD  models  in  the mass  range  0.17-0.41 \msun  were
evolved down to very low surface luminosities starting from physically
sound  initial models  generated by  abstracting mass  from a  1 \msun
model at appropriate  stages of its red giant  branch evolution.  Most
importantly,  gravitational settling,  chemical and  thermal diffusion
have been considered in these new calculations and model evolution was
computed  in  a self-consistent  way  with  the  evolution of  element
abundances  as  predicted  by  diffusion processes.   Althaus  et  al.
(2000a) find that element diffusion strongly affects the structure and
evolution  of He  WDs,  giving  rise to  a  different cooling  history
depending on the stellar mass of the models.  In particular, diffusion
induces the occurrence of hydrogen shell flashes in He WDs with masses
ranging from $\approx$ 0.18 to  0.41 \msun, which is in sharp contrast
from the  situation when diffusion  is neglected.  In  connection with
the  further  evolution,   these  diffusion-induced  flashes  lead  to
hydrogen envelopes thin enough so as to prevent stable nuclear burning
from   being  an   important   energy  source   at  advanced   cooling
stages.\footnote{  Since  Althaus  et  al. (2000a,b)  did  not  invoke
additional mass transfer when models return to the red giant region as
a  result of  thermonuclear flashes,  the surviving  hydrogen envelope
masses  after flash  episodes  are upper  limits.}  This implies  much
shorter  cooling ages  than in  the case  when diffusion  is neglected
(Driebe et  al.  1998).  On the  contrary, for masses  lower than 0.18
\msun (as  it is  the case  for the WD  companion to  PSR J1012+5307),
nuclear  burning is  dominant even  in the  presence of  diffusion and
cooling ages resemble those derived from models without diffusion.
 
The  evolutionary calculations  mentioned in  the  preceding paragraph
have  important  implications  when   attempt  is  made  in  comparing
theoretical  predictions on  the WD  evolution with  expectations from
millisecond pulsars, such as those  mentioned earlier.  As a matter of
fact, these  evolutionary models predict  that as a result  of element
diffusion, evolution  is accelerated  to such an  extent that  the age
discrepancy   between    the   PSR   B1855+09    components   vanishes
completely. More specifically, these models  lead to an age of 4 $\pm$
2 Gyr for the WD component in good agreement with the spin-down age of
the pulsar. As mentioned,  there also exists other millisecond pulsars
with WD companions  for which cooling and spin-down  ages appear to be
discrepant when  the WD  age is assessed  from evolutionary  models in
which diffusion is  not considered.  This is the  case for the systems
involving PSR J0034-053 and J1713+0747 (Hansen \& Phinney 1998b) which
most  likely have  very cool  He WD  companions. On  the contrary,
evolutionary  calculations including  element diffusion  give  ages in
good  agreement  with  the  ages  of  such  pulsars  (Althaus  et  al.
2000a). Thus, age discrepancies  between the predictions of standard
evolutionary  models  and  recent  observational data  of  millisecond
pulsar systems appear  to be the result of  ignoring element diffusion
in such evolutionary calculations.
  
According to the Althaus et al.  (2000a) evolutionary calculations, He
WDs with stellar masses greater  than $\approx$ 0.18 \msun could reach
very low  \teff stages well within  the age of the  universe.  In this
connection,  the  WD LSH  3250  is  particularly  noteworthy.  With  a
surface  luminosity  of  \ll_lsun=   -4.57  $  \pm$  0.04  (Harris  et
al. 1999), this places it  amongst the lowest luminosity known for any
WD.  It  is highly likely  that this WD  is characterized by  a helium
core and  a very low  \teff value (Harris  et al. 1999) such  that its
emergent spectrum  would be dominated  by collision-induced absorption
(CIA)  from molecular  hydrogen.  Such  a strong  molecular absorption
causes cool  WDs to become bluer  as they age (Hansen  1998; Saumon \&
Jacobson 1999 and  Rohrmann 2000; see also Saumon et  al.  1994 in the
context  of  low  mass  stars).  WD evolutionary  sequences  based  on
detailed  radiative  transfer calculations  appropriate  for very  old
carbon-oxygen WDs have recently  been presented by Hansen (1998, 1999)
and Salaris et al. (2000).

If cool  low-mass He WDs  are actually characterized by  short cooling
ages, as  claimed by Althaus et  al., then many of  them could present
blue colours  as a result of  strong CIA from  molecular hydrogen.  In
light of  these concerns, we judge  it to be  worthwhile to re-examine
the ages of  cool He WDs and  at the same time to  provide colours and
magnitudes  for  these  stars   in  a  self-consistent  way  with  the
predictions of  stellar evolution and element diffusion.   This is the
main  aim of  the present  work,  and to  our knowledge  such kind  of
calculations  have  never been  performed.   The present  calculations
improve over those presented in Althaus et al (2000a) in the fact that
here appropriate outer boundary  conditions for the cooling models are
derived on the basis  of detailed non-gray model atmospheres (Rohrmann
2000).  As far as cool WD ages are concerned, this is a very important
point because  WD cooling  is very sensitive  to the treatment  of the
outer boundary conditions. Another  motivation for the present work is
the determination of a theoretical luminosity function for He WDs from
our new cooling curves.   Details about our evolutionary code, initial
models and diffusion treatment are  briefly described in Section 2. In
that  section, we  also detail  the main  characteristic of  our model
atmospheres.  Results are presented  in Section  3; in  particular the
evolution  of our  models  in colour-magnitude  diagrams is  analysed.
Finally, Section 4 is devoted to making some concluding remarks.

\section{Computational details}

As  stated in the  introduction, one  of the  aim of  this work  is to
provide a grid  of ages, colours and magnitudes  for WD cooling models
appropriate for the study of old He WDs. This set of models covers
an wide range of  stellar masses as  expected for such  objects.  In
particular,  the evolution  of models  with stellar  masses  of 0.406,
0.360, 0.327,  0.292, 0.242, 0.196  and 0.169 \msun has  been followed
from the end of mass loss episodes during the pre-WD evolution down to
very  low  surface  luminosities.   WD  evolution  is  computed  in  a
self-consistent way  with the expectations from  element diffusion, as
done  in   Althaus  et  al.    (2000a,b);  but  in  contrast   to  the
gray-atmosphere approximation assumed in that work, we consider here a
detailed  treatment  of  the  atmosphere  which enable  us  to  obtain
accurate outer boundary conditions for our cooling models and colour
indices as well. 

Attention  is  focused  mainly  on  the low  \teff  regime,  where  WD
evolution is markedly dependent  on the treatment  of the  very outer
layers. When models reach  such stages of evolution, element diffusion
have  caused the  bulk of  hydrogen to  float and  helium  and heavier
elements  to sink  from outer  layers (see  Althaus et  al.  2000a for
details). This implies almost pure hydrogen atmospheres and justifies
our use of atmosphere models for zero metallicity (see below).

In  what  follows, we  briefly  comment  on  the initial  models,  our
atmospheric  treatment   and  the   input  physics  included   in  our
evolutionary code.

\subsection{Initial models}

Reliable initial models have  been obtained by simply abstracting mass
from a 1  \msun \ model at appropriate stages of  the red giant branch
evolution (see also Iben \& Tutukov 1986 and Driebe et al.  1998).  In
this way, we  were able to generate initial He  WD models with stellar
masses of  0.406, 0.360, 0.327,  0.292, 0.242, 0.196 and  0.169 \msun.
It  is worth  mentioning  that the  resulting  envelopes and  hydrogen
surface abundance of  these initial models are in  good agreement with
those quoted by Driebe et  al (1998).  We follow the further evolution
of such  initial configurations  consistently with the  predictions of
element diffusion.   As the stars enter the  cooling branch, diffusion
processes begin to play an important role.  First, they cause hydrogen
to  float and other  elements to  sink, giving  rise to  pure hydrogen
outer layers.  Simultaneously,  chemical diffusion makes some hydrogen
move  inwards  to  hotter   layers,  favouring  the  occurrence  of  a
thermonuclear flash.  Short after  the flash begins, the star suddenly
increases its radius and developes an outer convection zone which gets
deep  enough  and  reaches  helium-rich  layers,  thus  modifying  the
composition of the  outer layers. Over a time-scale of  the order of a
few hundred  years the star evolves  back to the red  giant region and
then  finally  to the  cooling  branch.  At  this stage,  evolutionary
time-scale gets longer and diffusion begins to be important again with
the result that the outer layers become made up of only hydrogen. This
in sharp  contrast with the case  in which diffusion  is neglected. In
that case, apart  from convection, no other physical  agent is able to
drive  hydrogen to  the surface  and therefore  the final  outer layer
chemical composition is fixed by the last episode of convective mixing
(made up by hydrogen and helium in comparable proportions). Should the
hydrogen  envelope  left after  the  flash  episode  is thick  enough,
another thermonuclear  flash will be triggered  by chemical diffusion.
This  sequence  of events  will  finish  once  the remaining  hydrogen
envelope is thin  enough for hydrogen burning to  be negligible.  Thus
the amount  of hydrogen remaining  after hydrogen flashes  have ceased
results markedly lower when diffusion is allowed to operate. This will
produce  not only  a different  cooling  history as  mentioned in  the
introduction but also noticeable changes in the structure of the star.
In this regard, surfaces  gravities result very different if diffusion
is considered. For  models with $M_* < $ 0.18  \msun for instance, the
surface  gravity is  reduced by  almost  80 per  cent as  a result  of
element diffusion (see Althaus et al.  2000a for details).

After  flash episodes  have ceased  and  when models  reach the  final
cooling  branch,   we  compute  the  subsequent   evolution  by
considering a detailed treatment  of the atmosphere (see below). Here,
evolution becomes slow enough for the purity of the outer layers to be
established by  diffusion processes.\footnote{However, we  should note
that, as a result of the long diffusion time-scales at the base of the
outer convection zone, metals  accreted from interstellar medium could
be mantained in the  outer layers of very cool He WDs  for a long time
(see Althaus \& Benvenuto 2000)}

\subsection{Model Atmosphere}

For a proper treatment of  the cooling behaviour of He WDs we
have  calculated  the   evolution  of  our  He  WD   models  in  a
self-consistent way  with the  predictions of detailed  non-gray model
atmospheres,  which  are  described  at  length  in  Rohrmann  (2000).
Recently, the  atmospheric code described in Rohrmann  (2000) has been
adapted  for  treating  mixtures  of  hydrogen and  helium.   Here  we
restrict ourselves to a few brief  comments and we refer the reader to
that  work for  details  and  comparisons of  such  models with  other
relevant atmospheric calculations in the literature.

To begin  with, our non-gray  model atmospheres are  constructed under
the  assumption of constant  gravity, local  thermodynamic equilibrium
and plane-parallel  geometry and  include hydrogen and  helium species
(zero metallicity).   Energy transfer by radiation  and convection are
taken into account and the  resulting equations are solved by means of
a standard linearization technique. More specifically, the equation of
radiative  transfer (formulated  in  terms of  the variable  Eddington
factors,  see  Auer  \&  Mihalas  1970) and  constant  flux  condition
(radiative  and convective contributions)  are solved  by means  of an
iterative  procedure over linearized  equations for  a set  of optical
depth points  and a  frequency mesh.  In  the interests  of minimising
computing  times, we  consider  a partial  linearization procedure  in
terms  of the  temperature alone  (Gustafsson 1971  and  Gustafsson \&
Nissen 1972).   Only the Planck  function and the convective  flux are
linearized  and  temperature  corrections  at  each  depth  point  are
obtained  by means  of the  Rybicki procedure  in which  the resulting
system  of linear equations  are ordered  not in  depth blocks  but in
frequency blocks (see Gustaffson \& Nissen 1972).

Constitutive physics  of our  model atmospheres is  based on  an ideal
equation  of  state\footnote{Non-ideal  effects 
become  important in hydrogen model atmospheres for
\teff $<$  2500K; see  Saumon \& Jacobson  (1999).} and  the following
species  have   been  considered:  H,  H$_2$,   e$^-$,  H$^-$,  H$^+$,
H$_2$$^+$,   H$_3$$^+$,  He,  He$^+$   and  He$^{++}$.   Although  the
abundances  of species  such as  H$_2$$^+$ and  H$_3$$^+$  are usually
negligible,  their presence  affects significantly  the  absorption and
emission of radiation (see  Saumon et al.  1994).  Partition functions
of H and H$_2$; H$_2$$^+$  and H$_3$$^+$ are from Irwin (1981), Sauval
\&  Tatum (1984)  and  Neale \&  Tennyson  (1995), respectively.   All
relevant bound-free,  free-free and scattering  processes contributing
to  opacity have  been  included  in our  calculations.  At low  \teff
values, CIA by molecular hydrogen  due to collisions with H$_2$ and He
represents a major source of opacity in the infrared and dominates the
shape of the  emergent spectrum at low \teff.   Here, we have included
calculations of  H$_2$-H$_2$ and  H$_2$-He  CIA cross
sections by Borysow, Jorgensen \& Zheng (1997).  Convection is treated
within the  formalism of  the mixing length  (ML2 version) and  it has
been included self-consistently in  the equation of energy conservation
and in  the linearization procedure.  It is worth mentioning  that for
\teff values below 8000K,  the temperature profile becomes insensitive
to the adopted parametrization of the mixing-length theory (Bergeron,
Wesemael  \&  Fontaine  1992).   Broadband colour  indices  have  been
calculated using  the optical $BVRI$  and infrared $JHK$  passbands of
Bessell  (1990)  and  Bessell  \&  Brett  (1988),  respectively,  with
callibration constants from Bergeron, Ruiz \& Leggett (1997).

We  note that,  as reported  by Saumon  et al.   (1994)  and Bergeron,
Saumon \& Wesemael (1995),  numerical instabilities resulting from the
competition between  CIA by H$_2$  and H$^{-}$ opacity develop  in the
\teff  regime  from  $\approx$  3500-5000K depending  on  the  surface
gravity. Such  instabilities were overcome by  the procedure suggested
by these  authors. Finally,  starting pressure and  temperature values
needed to  integrate the  envelope equations are  given at  an optical
depth $\tau$ of $\approx$ 25. At this point the diffusion approximation
for  the  radiative transfer  can  be assumed  to  be  valid and  the
equations of stellar structure in the envelope, at the bottom of which
we specify the outer boundary  conditions to the evolutive models, can
be integrated using Rosseland mean opacities.

\begin{figure}
\epsfxsize=240pt \epsfbox[19 350 578 795]{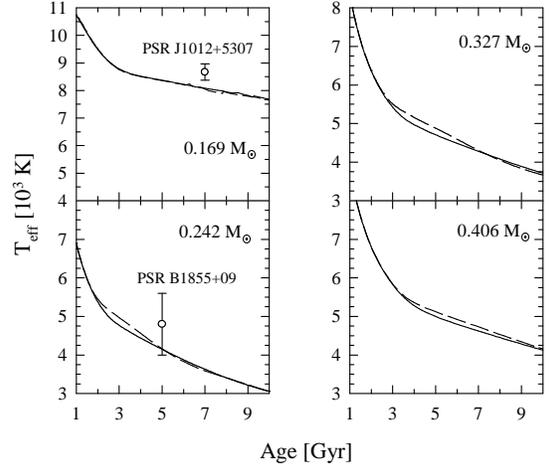}
\caption {Effective temperature  as a function of age for
some selected values  of the stellar mass.  Solid  lines correspond to
the  case  when  a  detailed  non-gray atmosphere  is  considered  for
deriving the outer boundary conditions  to the WD evolution and dashed
line  to  the  situation  when  the  standard  gray  approximation  is
used. Both  sets of calculations take into  account element diffusion.
The observational data for the WD companion to the millisecond pulsars
PSR  B1855+09 and  PSR J1012+5307  are included.   Note that  for more
massive models  and 4000K $\lesssim T_{\rm eff} \lesssim$  5500K  cooling 
ages become  smaller in the case  of  a non-gray treatment
for the atmosphere.}
\end{figure}

\subsection{Evolutionary code and diffusion treatment}

The evolutionary  calculations presented  in this work  were performed
using  the evolutionary  code described  in our  previous works  on WD
evolution (see, e.g, Althaus \& Benvenuto 1997, 2000 and 
Benvenuto \& Althaus 1998).  Shortly,
our code includes a detailed  and updated physical description such as
OPAL  radiative  opacities (Iglesias  \&  Rogers  1996) and  molecular
opacities (Alexander \& Ferguson 1994);  the equation of state for the
low  density  regime  is  an  updated  version of  that  of  Magni  \&
Mazzitelli  (1979), whilst  for the  high density  regime  we consider
ionic   contributions,  Coulomb  interactions,   partially  degenerate
electrons,  and electron  exchange and  Thomas-Fermi  contributions at
finite temperature.   Conductive opacities and  the various mechanisms
of  neutrinos emission  are taken  from  the formulation  of Itoh  and
collaborators.  Hydrogen burning is  considered via a complete network
of  thermonuclear reaction  rates corresponding  to  the proton-proton
chain  and the CNO  bi-cycle.  The  chemical evolution  resulting from
element   diffusion   has  also   been   considered.   The   diffusion
calculations  are based  on the  multicomponent treatment  of  the gas
developed by  Burgers (1969) and gravitational  settling, chemical and
thermal diffusion have been taken  into account. We mention that the WD
evolution is calculated in  a self-consistent way with the predictions
of element  diffusion. In addition, radiative  opacities are calculated
for metallicities as given by the varying abundances (more details are
given in Althaus et al. 2000a and Althaus \& Benvenuto 2000).

\begin{figure}
\epsfxsize=240pt \epsfbox[19 300 578 795]{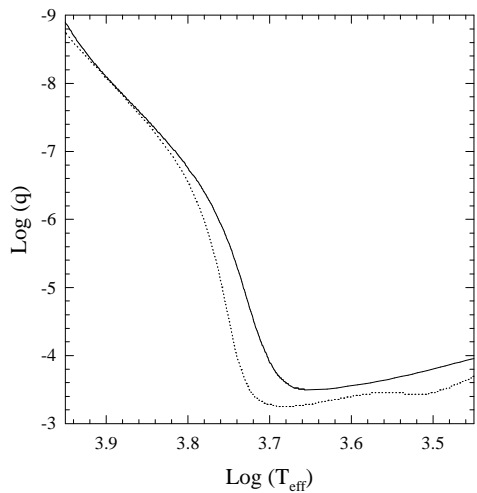}
\caption {The location of the base of the outer convection zone
in terms of the outer mass fraction $q$ as a function of the
effective temperature for 0.242  M$_{\odot}$ He WD models. 
Solid and dotted lines correspond,
respectively, to the  non-gray and gray treatment for the atmosphere.
Note that the outer convection zone is shallower when account is made of
the detailed treatment for the atmosphere.}
\end{figure}

\section{Results} \label{sec:results}

We begin by examining the  evolutionary ages for some selected stellar
masses.  They are illustrated in Fig.  1 in which the \teff versus age
relation is depicted  together with the observational data  for the WD
companions  to the  millisecond  pulsars PSR  J1012+5307 and  B1855+09
(Callanan et al.   1998 and van Kerkwijk et  al.  2000, respectively).
Only  the  evolution corresponding  to  the  final  cooling branch  is
depicted  in the figure.   For stellar  masses greater  than $\approx$
0.18 \msun,\ models  experience diffusion-induced hydrogen flashes and
they are left with small hydrogen mass and little nuclear burning with
the consequent  result that cooling ages  become substantially smaller
as compared with  the situation for less massive  models, which do not
suffer from flash episodes (see  Althaus et al.  2000a). Note the good
agreement with the observational  predictions for both of the pulsars.
In particular,  for the PSR B1855+09 companion,  models with diffusion
predict an age of 4 $\pm$ 2 Gyr in good agreement with the pulsar age.
It is clear  that element diffusion is an  important ingredient in the
evolution of  He WDs that must  be taken into account  when attempt is
made in comparing theoretical predictions with observational data.

\begin{figure}
\epsfxsize=240pt \epsfbox[19 350 578 795]{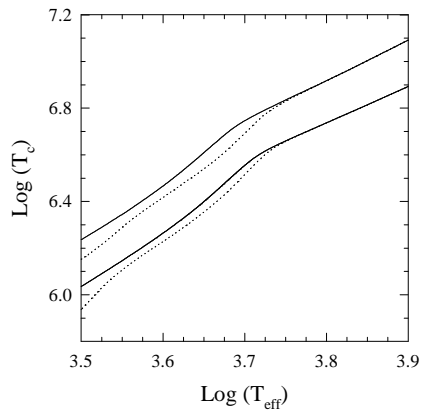}
\caption {Central temperature as a function of effective temperature  
for 0.242 and 0.406 M$_{\odot}$ He WD models (upper and lower curves,
respectively). For each stellar mass, solid and dotted lines correspond,
respectively, to the non- gray and gray treatment for the atmosphere.} 
\end{figure}

Note  that  differences  in the  cooling  of  He  WDs arise  from  the
employment of detailed  model atmospheres. Specifically, the inclusion
of proper outer  boundary conditions may decrease the  cooling ages up
to 1  Gyr in  the range 4000K  $\lesssim$ \teff $\lesssim$  5500K.  At
such  stages of  evolution, the  central temperature  becomes strongly
tied to the  temperature stratification of the outer  layers. In fact,
when  convection  reaches  the   domain  of  degeneracy,  the  central
temperature drops  substantially and the star has  initially an excess
of internal energy  to be radiated, thus giving  rise to a lengthening
of the  evolutionary times  during that epoch  of evolution, as  it is
borne out by  the change of slope in the cooling  curves shown in Fig.
1.  At advanced stages of evolution, the thermal stratification of the
envelope is affected by the use  of non-gray atmospheres in such a way
that the  location of  the maximum  depth reached by  the base  of the
outer  convection zone  is  markedly shallower  as  compared with  the
prediction of the standard gray treatment for the atmosphere (see also
Bergeron et al.  1997). This can be appreciated in Fig.  2 in which we
show the evolution  of the base of the outer  convection zone in terms
of \teff for the 0.242  \msun model.  Clearly, when \teff is decreased
below  7000K,  the  non-gray  treatment  predicts  a  shallower  outer
convection zone  (this behaviour is essentially  the same irrespective
of the  stellar mass of the  model).  As a  result, convection reaches
the  degenerate  core  at  lower  \teff  values  in  such  models  and
accordingly the drop in the central temperature takes place later than
for the gray models, as shown in Fig.  3.  This behaviour explains the
trend of the cooling curves in both set of calculations.

\begin{figure}
\epsfxsize=240pt \epsfbox[19 260 578 795]{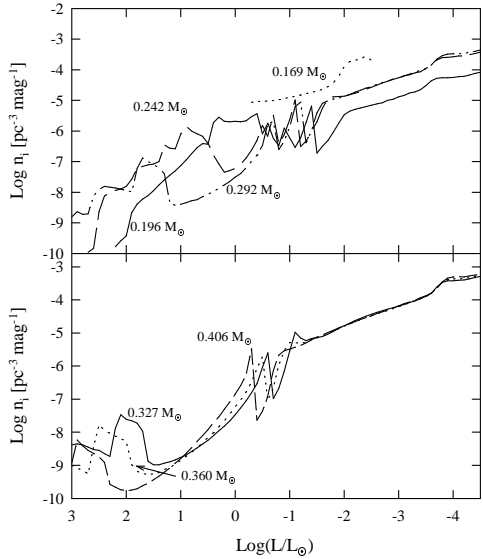}
\caption{ Single  luminosity  functions versus  surface
luminosity for our  He  WDs models.   The  spikes at  intermediate
luminosities correspond to the evolutionary phases before the onset of
flash episodes  and the  bumps at high  luminosities reflect  the slow
rate of evolution before the  models enter their cooling branch.  Less
massive models (top panel)  experience various flash episodes, so they
reach the  blue part of the  HR diagram many  times.  Accordingly, the
resulting luminosity functions show more extended bumps.}
     
\end{figure}

Another  interesting  issue  of  our  evolutionary  calculations  also
related to the  extent of the outer convection  zone is the following.
As noted by Althaus et  al.  (2000a), the hydrogen envelope left after
diffusion-induced flash episodes may be thin enough that convection is
able to mix it with the underlying helium layers at advanced stages of
evolution.  In fact, these authors find that over a time interval of 2
Gyr, models  with masses of  $\approx$ 0.2 \msun are  characterized by
outer layers  made up  by hydrogen  and helium by  the time  they have
cooled down to \teff $\approx$ 5000K.  As mentioned (see Fig. 2) , the
use of detailed model  atmospheres to derive outer boundary conditions
for evolving  WDs gives  rise to shallower  outer convection  zones as
compared with the gray treatment considered in Althaus et al. (2000a).
As a result,  no mixing episodes occur at all and  the envelope of our
He WD models  remains as pure hydrogen down to the  end of the cooling
sequence.

From the cooling curves, we  can estimate the predicted number density
of He WDs as a function  of luminosity. To this end we have translated
the  cooling curves  for our  set of  models into  a  total luminosity
function according to the relation given by

\begin{equation}
\Phi= \sum_i{n_i},
\label{funlumi}
\end{equation}

\noindent with 

\begin{equation}
n_i= \frac{k_i \Delta t}{V}.
\end{equation} 

\noindent  Here, $k_i$  is  the birthrate  of  He WDs  of a  given
stellar mass, the summation of which is normalized to the birthrate at
which He  WDs are produced  (0.14 $yr^{-1}$; see Iben,  Tutukov \&
Yungelson 1997).   $n_i$ is the  number density of He  WDs (single
luminosity  function)  of  a  given  mass in  the  magnitude  interval
M$_b-0.5$  to M$_b+0.5$,  $\Delta t$  is the  time interval  the model
requires to  evolve from M$_b-0.5$  to M$_b+0.5$ magnitude and  $V$ is
the  volume of the  Galactic disk  ($3 \times  10^{11} pc^3$).  On the
basis of these assumptions,  the resulting single luminosity functions
for our models  are displayed in Fig. 4.  A  striking feature shown by
this  figure  are  the  spikes  at  intermediate  luminosities,  which
correspond  to the  evolutionary phases  prior to  the onset  of flash
episodes where evolution slows down. At high luminosities, each He
WD  theoretical  distribution exhibits  a  bump.   These  bumps are  a
consequence  of the  slow rate  of evolution  before the  models enter
their cooling  branch.  To understand  the behaviour of $n_i$  at high
luminosities,  it  should  be  kept  in  mind  that  as  a  result  of
thermonuclear flashes, models reach  the high surface luminosity phase
more than once.  Less  massive models experiences many flash episodes,
thus their resulting luminosity functions show more extended bumps.
    
\begin{figure}
\epsfxsize=240pt \epsfbox[19 340 578 795]{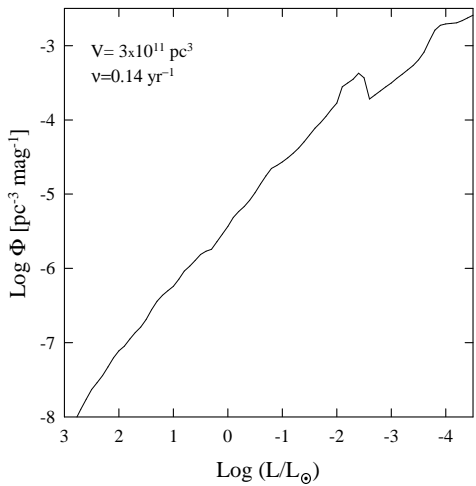}
\caption{Total luminosity function  (resulting from the
superposition of  the single number-density of He  WD given in
Figure 4)  versus surface  luminosity for our  He WD  models.  The
bump  at $log(L/L_{\odot})  \approx -2$  is due  to the  slow  rate of
evolution  characterizing our  lowest mass  model, which  cannot reach
lower luminosities within a Hubble time. Such low mass He WDs are
thus expected not to contribute to the luminosity function at very low 
luminosities. For details, see text.}

\end{figure}

The total luminosity function,  $\Phi$, results from the superposition
of the single number density of He WDs and it is illustrated by Fig.5.
Note that over  most of the given surface  luminosity range, the total
luminosity  function exhibits a  rather monotonously  increasing trend
down  to \ll_lsun $\approx$  - 2.5.   Little of  the behaviour  of the
individual  luminosity  function remains.   This  is expected  because
flash episodes  occur at  different surface luminosities  depending on
the stellar mass with the  result that spikes cancel amongst them out.
At \ll_lsun  $\approx$ -  2.5, the luminosity  function shows  a local
maximum  as  a result  of  the contribution  of  models  which do  not
experience hydrogen shell flashes.  Indeed, as mentioned earlier, such
models are characterized  by stable nuclear burning even  at low \teff
values,   with  the   consequent  result   that  their   evolution  is
considerably slowed down. More specifically, at such a luminosity, the
0.169 \msun model  is characterized by an age of  $\approx$ 14 Gyr; so
it is  expected not to contribute to  the number density of  He WDs at
lower  luminosities  within the  age  of  the  universe.  At  \ll_lsun
$\approx$  -  4, the  luminosity  function  shows  a small  bump.   As
explained earlier, when convection  reaches the degenerate core, there
is  initially an  excess of  energy  to get  rid of  and the  cooling
process  is  slowed  down, thus  leading  to  the  small bump  in  the
luminosity function at low  stellar luminosities. 
Finally, we should note that
we have considered  single He WDs, that is  we have neglected possible
merger of those He WDs  formed in binary systems with small separation
for gravitational wave radiation to lead to a merger of the components
on time scales shorter than the Hubble time.  Accordingly, care should
be  taken at  comparing our  theoretical luminosity  functions  at low
luminosities with observations.

\begin{table*}
\centering
\begin{minipage}{120mm}
\caption{Selected stages for 0.169, 0.242, 0.327 and 0.406 \msun 
 He WD models}
\begin{tabular}{@{}cccccccccccc@{}}
\hline
$M_*/{\rm M_{\odot}}$ & 
\teff &  $Log (g)$ &  $Age$ (Gyr) &B-V &
V-R & V-K & R-I & J-H & H-K & BC & M$_V$ \\
\hline
 0.169 & 10000 & 6.0621 &  1.56 & -0.01 &  0.04 &  0.03 &  0.06 &  0.02 & -0.06 & -0.34 &   8.70 \\
 '' &  9500 & 6.1624 &  1.99 &  0.04 &  0.07 &  0.15 &  0.08 &  0.04 & -0.06 & -0.29 &   9.11 \\
 '' &  9000 & 6.2555 &  2.60 &  0.10 &  0.11 &  0.30 &  0.11 &  0.07 & -0.05 & -0.25 &   9.55 \\
 '' &  8500 & 6.3499 &  4.11 &  0.16 &  0.14 &  0.46 &  0.14 &  0.10 & -0.04 & -0.23 &  10.01 \\
 '' &  8000 & 6.4256 &  7.63 &  0.22 &  0.18 &  0.64 &  0.18 &  0.14 & -0.02 & -0.21 &  10.44 \\
 '' &  7500 & 6.4796 & 11.45 &  0.28 &  0.22 &  0.82 &  0.22 &  0.18 & -0.01 & -0.19 &  10.84 \\
 '' & 7050 &  6.5265 &  16.52 &  0.34 &  0.25 &  1.00 &  0.25 &  0.21 &  0.00 & -0.17 &  11.21 \\
\hline
\\
 0.242 & 10000 & 6.9458 &  0.32 &  0.03 &  0.07 &  0.08 &  0.07 &  0.03 & -0.06 & -0.37 &  10.54 \\
 '' &  9500 & 6.9616 &  0.37 &  0.08 &  0.09 &  0.20 &  0.09 &  0.05 & -0.05 & -0.32 &  10.76 \\
 '' &  9000 & 6.9780 &  0.44 &  0.13 &  0.12 &  0.34 &  0.12 &  0.08 & -0.04 & -0.28 &  11.00 \\
 '' &  8500 & 6.9945 &  0.53 &  0.18 &  0.15 &  0.49 &  0.15 &  0.11 & -0.03 & -0.25 &  11.26 \\
 '' &  8000 & 7.0119 &  0.64 &  0.23 &  0.19 &  0.65 &  0.18 &  0.14 & -0.02 & -0.22 &  11.53 \\
 '' &  7500 & 7.0291 &  0.78 &  0.29 &  0.22 &  0.83 &  0.22 &  0.18 & -0.01 & -0.20 &  11.83 \\
 '' &  7000 & 7.0464 &  0.96 &  0.35 &  0.26 &  1.02 &  0.26 &  0.22 &  0.01 & -0.18 &  12.15 \\
 '' &  6500 & 7.0637 &  1.19 &  0.43 &  0.30 &  1.23 &  0.30 &  0.26 &  0.03 & -0.16 &  12.50 \\
 '' &  6000 & 7.0818 &  1.48 &  0.52 &  0.35 &  1.48 &  0.36 &  0.30 &  0.06 & -0.17 &  12.90 \\
 '' &  5500 & 7.1044 &  1.88 &  0.64 &  0.42 &  1.76 &  0.42 &  0.34 &  0.09 & -0.21 &  13.38 \\
 '' &  5000 & 7.1423 &  2.50 &  0.77 &  0.50 &  2.11 &  0.50 &  0.39 &  0.13 & -0.31 &  13.99 \\
 '' &  4500 & 7.1779 &  3.80 &  0.91 &  0.59 &  2.36 &  0.59 &  0.38 &  0.09 & -0.45 &  14.68 \\
 '' &  4000 & 7.1974 &  5.54 &  1.04 &  0.67 &  2.15 &  0.67 &  0.10 & -0.01 & -0.51 &  15.29 \\ 
 '' &  3500 & 7.2082 &  7.57 &  1.15 &  0.73 &  1.42 &  0.70 & -0.18 & -0.19 & -0.37 &  15.76 \\
 '' &  3000 & 7.2149 & 10.38 &  1.25 &  0.74 &  0.39 &  0.56 & -0.30 & -0.32 & -0.07 &  16.14 \\
 '' &  2500 & 7.2207 & 15.20 &  1.36 &  0.65 & -0.79 &  0.00 & -0.21 & -0.48 &  0.34 &  16.54 \\
\hline
\\
 0.327 & 10000 & 7.3138 &  0.57 &  0.05 &  0.07 &  0.10 &  0.07 &  0.03 & -0.06 & -0.38 &  11.15\\
 '' &  9500 & 7.3240 &  0.68 &  0.09 &  0.10 &  0.22 &  0.09 &  0.05 & -0.05 & -0.34 &  11.35\\
 '' &  9000 & 7.3339 &  0.80 &  0.14 &  0.13 &  0.36 &  0.12 &  0.08 & -0.04 & -0.30 &  11.57\\
 '' &  8500 & 7.3433 &  0.94 &  0.19 &  0.16 &  0.50 &  0.15 &  0.11 & -0.03 & -0.26 &  11.81\\
 '' &  8000 & 7.3524 &  1.12 &  0.24 &  0.19 &  0.66 &  0.18 &  0.14 & -0.02 & -0.23 &  12.07\\
 '' &  7500 & 7.3613 &  1.33 &  0.29 &  0.22 &  0.83 &  0.22 &  0.18 & -0.01 & -0.20 &  12.34\\
 '' &  7000 & 7.3702 &  1.58 &  0.36 &  0.26 &  1.02 &  0.26 &  0.22 &  0.01 & -0.18 &  12.63\\
 '' &  6500 & 7.3794 &  1.90 &  0.43 &  0.30 &  1.23 &  0.30 &  0.26 &  0.03 & -0.16 &  12.96\\
 '' &  6000 & 7.3899 &  2.31 &  0.53 &  0.36 &  1.47 &  0.36 &  0.29 &  0.06 & -0.16 &  13.34\\
 '' &  5500 & 7.4041 &  2.89 &  0.65 &  0.42 &  1.76 &  0.43 &  0.33 &  0.10 & -0.21 &  13.80\\
 '' &  5000 & 7.4258 &  3.91 &  0.77 &  0.50 &  2.10 &  0.50 &  0.39 &  0.12 & -0.31 &  14.37\\
 '' &  4500 & 7.4451 &  5.95 &  0.91 &  0.59 &  2.30 &  0.59 &  0.35 &  0.07 & -0.44 &  15.00\\
 '' &  4000 & 7.4559 &  8.41 &  1.03 &  0.67 &  2.02 &  0.66 &  0.04 & -0.03 & -0.47 &  15.57 \\
 '' &  3500 & 7.4622 & 11.39 &  1.13 &  0.72 &  1.27 &  0.68 & -0.21 & -0.20 & -0.32 &  16.02 \\
 '' &  3000 & 7.4665 & 15.67 &  1.23 &  0.72 &  0.26 &  0.51 & -0.30 & -0.32 & -0.01 &  16.39 \\
 '' &  2500 & 7.4701 & 23.95 &  1.33 &  0.61 & -0.96 & -0.10 & -0.20 & -0.51 &  0.39 &  16.79 \\
\hline
\\
 0.406 & 10000 & 7.5664 &  0.71 &  0.06 &  0.08 &  0.12 &  0.07 &  0.03 & -0.06 & -0.39 &  11.55 \\
 '' &  9500 & 7.5722 &  0.82 &  0.10 &  0.11 &  0.24 &  0.10 &  0.05 & -0.05 & -0.35 &  11.75\\
 '' &  9000 & 7.5783 &  0.95 &  0.14 &  0.13 &  0.37 &  0.12 &  0.08 & -0.04 & -0.31 &  11.96\\
 '' &  8500 & 7.5838 &  1.11 &  0.19 &  0.16 &  0.51 &  0.15 &  0.11 & -0.03 & -0.27 &  12.18\\
 '' &  8000 & 7.5897 &  1.30 &  0.24 &  0.19 &  0.67 &  0.18 &  0.14 & -0.02 & -0.24 &  12.43\\
 '' &  7500 & 7.5952 &  1.55 &  0.29 &  0.22 &  0.84 &  0.22 &  0.18 & -0.01 & -0.21 &  12.69\\
 '' &  7000 & 7.6010 &  1.85 &  0.36 &  0.26 &  1.02 &  0.26 &  0.22 &  0.01 & -0.18 &  12.98\\
 '' &  6500 & 7.6072 &  2.25 &  0.44 &  0.30 &  1.23 &  0.30 &  0.25 &  0.03 & -0.16 &  13.29\\
 '' &  6000 & 7.6139 &  2.77 &  0.53 &  0.36 &  1.47 &  0.36 &  0.29 &  0.06 & -0.16 &  13.66\\
 '' &  5500 & 7.6238 &  3.52 &  0.65 &  0.42 &  1.75 &  0.42 &  0.33 &  0.10 & -0.21 &  14.11\\
 '' &  5000 & 7.6374 &  5.00 &  0.77 &  0.50 &  2.09 &  0.50 &  0.38 &  0.12 & -0.31 &  14.66\\
 '' &  4500 & 7.6482 &  7.71 &  0.91 &  0.59 &  2.25 &  0.59 &  0.33 &  0.06 & -0.43 &  15.27\\
 '' &  4000 & 7.6543 & 10.83 &  1.02 &  0.66 &  1.92 &  0.66 &  0.01 & -0.06 & -0.44 &  15.80\\
 '' &  3500 & 7.6580 & 14.68 &  1.12 &  0.71 &  1.15 &  0.67 & -0.22 & -0.22 & -0.28 &  16.23\\
 '' &  3000 & 7.6607 & 20.37 &  1.21 &  0.71 &  0.16 &  0.47 & -0.30 & -0.33 &  0.02 &  16.60\\
\\
\hline
\end{tabular}
\medskip


\end{minipage}
\end{table*}

The  evolution  of  the  emergent  flux  distribution  for  the  0.292
M$_{\odot}$ He WD  model at some selected \teff values  can be seen in
Fig.  6.   The results  of the non-gray  atmosphere (solid  lines) are
compared  with the blackbody  predictions (dashed  lines) at  the same
\teff value.  We want  to stress  again that by  the time  models have
reached the  \teff range considered  in Fig.6, element  diffusion have
caused  heavier  elements  than  hydrogen  to  sink  below  the  outer
layers.  So, our  evolving He  WD models  at such  advanced  stages of
evolution are  characterized by pure  hydrogen outer layers.   At high
temperatures, the  $H^-$ opacity is  dominant.  Since this  opacity is
almost independent  of frequency, the emergent flux  resembles that of
the blackbody  spectrum. The  main observation we  can make  from this
figure  is  that when  \teff  is  lowered  below 4000K,  the  emergent
spectrum becomes bluer.   This effect, reported also by  Saumon et al.
(1994) and  more  recently  by   Saumon  \&  Jacobson  (1999),  Hansen
(1998,1999), Rohrmann (2000) and  Salaris et al.  (2000) (particularly
in the context of carbon-oxygen WDs), is due to the strong CIA opacity
by  molecular  hydrogen  that  reduces  the infrared  flux  and  force
radiation to emerge at larger frequencies.  As a result, cool low-mass
He WDs  become bluer as  they age. Needless  to say, this  effect also
affects the colour-colour  and colour-magnitude diagrams. Indeed, most
colour indices show a pronounced turn-off at low \teff values and they
become bluer after reaching a  maximum.  This is nicely illustrated by
Fig. 7 in which the ($B-V$, $V-K$) two-colour diagram is shown for all
of our  He WD models together  with the observational data  for DA and
non-DA WDs according to Bergeron, Leggett \& Ruiz (2000). The peculiar
WD LHS 3250  analysed by Harris et al. (1999) is  also included in the
figure.  Note also that for \teff below 5000K, colours become markedly
bluer in this diagram.

\begin{figure}
\epsfxsize=240pt \epsfbox[19 130 578 795]{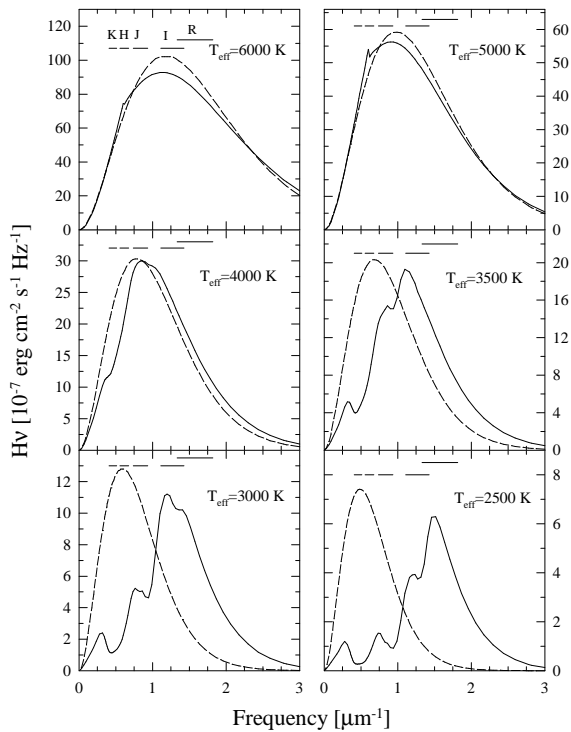}
\caption{ Emergent spectrum  for the  0.292 M$_{\odot}$
He WD  model at selected values of  the effective temperature
(as indicated within each figure).   Solid lines depict the results of
the  non-gray  atmosphere,  whilst  dashed  lines  correspond  to  the
blackbody  predictions at  the  same temperature.   Short solid  lines
indicate the  location of the  transmission functions for  the filters
$K,H,J,I$ and $R$.  Note that as effective temperature is lowered, the
spectrum  becomes bluer.   This is  a  direct consequence  of the  CIA
opacity which is dominant at low effective temperatures.}

\end{figure}

Now  we   analyse  the   evolution  of  our   He  WD  models   in  the
colour-magnitude diagram. In particular we  show in Figs.  8 and 9 the
behaviour of the  absolute visual magnitude $M_V$ for  our models as a
function  of the  color indices  $V-I$ and  $B-V$,  respectively.  The
observational  sample for  cool  DA (open  circles)  and non-DA  (open
triangles) WDs  of Bergeron  et al.  (2000)  are also included  in the
figure as well as the location  of the WD LHS 3250 according to Harris
et al. (1999). For the 0.169  \msun model, only the phase of low $M_V$
values is depicted because such a model would require exceedingly high
ages to evolve to lower luminosities. By contrast, more massive models
can reach high magnitudes within a Hubble time. Indeed, as can be seen
from the  figure, He WD models in  the mass range from  $\sim$ 0.18 to
0.3 M$_{\odot}$  can reach  the turn-off point  and become  blue again
within cooling times less than  15 Gyr. Specifically, such models have
cooling  ages of  6 -  9 Gyr  at the  turn-off, which  occurs  at $M_V
\approx$ 16.   Note that with further evolution,  they remain brighter
than  $M_V  \approx$  16.5.   The  results presented  here  raise  the
possibility that  many low-mass He WDs  could have had  time enough to
evolve beyond the  turn-off point and present blue  colours.  The cool
WD LHS  3250 analysed by Harris et  al. (1999) could be  an example of
such  WDs.  More  massive  He WDs  present  a similar  trend in  their
colour-magnitude  evolution  but the  ages  involved are  considerably
larger.  For  instance, the  0.406 \msun model  needs about 13  Gyr to
reach the turn-off in the $V-I$ colour index.  We would like to stress
again the fact that the  short cooling ages characterizing such models
are  due  to   the  role  played  by  element   diffusion  during  the
evolutionary phases  prior to those  computed here.  In  fact, element
diffusion leads to hydrogen shell flashes during which the mass of the
hydrogen envelope  is reduced to  such an extent that  nuclear burning
becomes  negligible on  the final  cooling track,  thus  implying much
shorter cooling  ages from the  situation when diffusion  is neglected
(such as in Driebe et  al. 1998). The colour-magnitude diagram for the
$B-V$ colour  index is  shown in  Fig. 9. Note  that the  $B-V$ colour
index does  not show a turn-off  at low \teff. We  should mention that
for \teff values above $\approx$  8000K, our predictions for the $B-V$
show discrepancies with those of Bergeron et al. (1995) because in our
calculations  we have not  considered the  effects of  line broadening
opacities.

\begin{figure}
\epsfxsize=240pt \epsfbox[19 240 578 795]{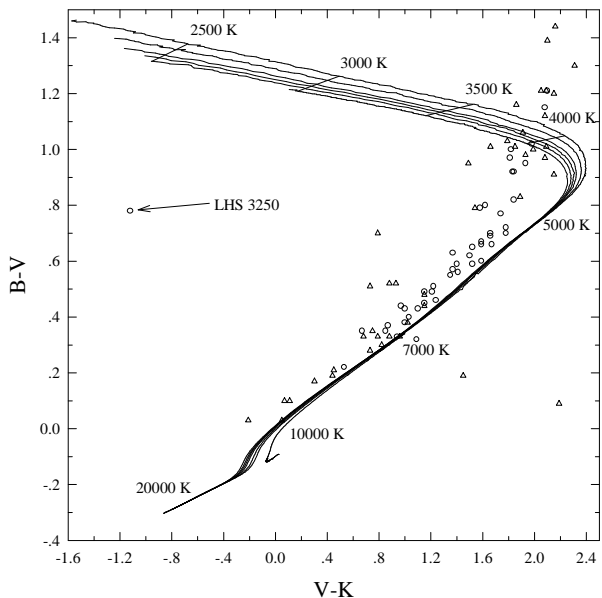}
\caption{($B-V$,  $V-K$) colour-colour  diagram  for our
He WD  models with  masses of (from  top to bottom)  0.196, 0.242,
0.292, 0.327,  0.360 and 0.406 M$_{\odot}$. Points  of equal effective
temperature  are  represented  by  thin  lines and  labeled  with  the
corresponding  values.  The  observational  sample for  cool DA  (open
circles) and non-DA  (open triangles) WDs of Bergeron  et al. (2000) are
also included in the figure as well as the location of the WD LHS 3250
according to Harris et al. (1999) determinations.}

\end{figure}

\begin{figure}
\epsfxsize=240pt \epsfbox[19 240 578 795]{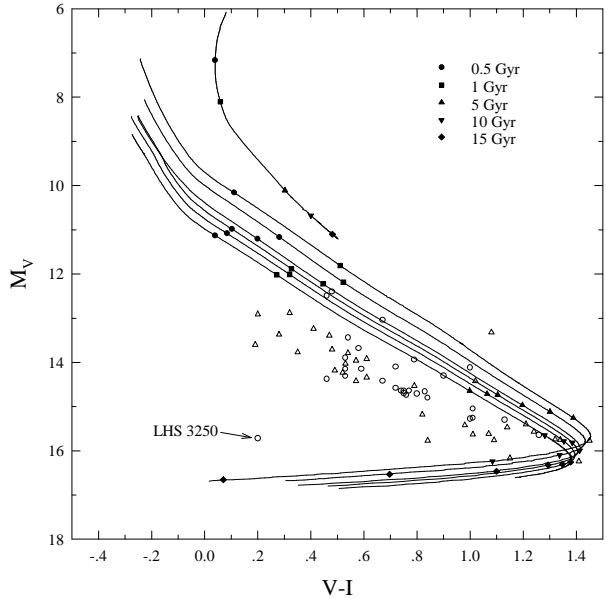}
\caption{Absolute visual magnitude $M_V$ in terms of the
colour index $V-I$ for our He WD models with masses of (from
top  to bottom)  0.169, 0.196,  0.242,  0.292, 0.327,  0.360 and  0.406
M$_{\odot}$. On  each curve, filled symbols indicate  cooling ages, as
explained  within the  figure. The  observational sample  for  cool DA
(open circles) and non-DA (open  triangles) WDs of Bergeron et al. (2000)
are also included in the figure as  well as the location of the WD LHS
3250 according to Harris et al. (1999) observations. Models with $M \geq$
0.196 M$_{\odot}$ exhibit a  pronounced turn-off at advanced stages of
evolution and then becomes  bluer with further evolution. According to
our evolutionary  calculations, He WDs with masses from $\sim 0.18$ 
to 0.3 M$_{\odot}$ could evolve beyond the turn-off well within 15 Gyr.}
\end{figure}

\begin{figure}
\epsfxsize=240pt \epsfbox[19 240 578 770]{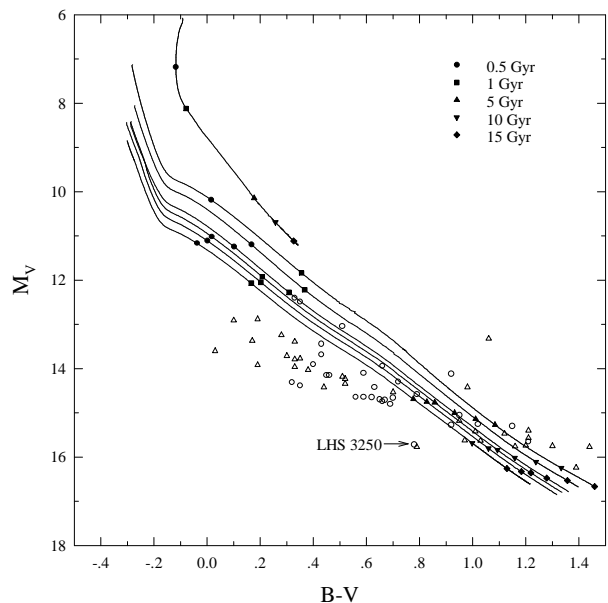}
\caption{Same  as Figure 8 but  for the  colour index
$B-V$.  Unlike the $V-I$, the $B-V$ colour index does not show a turn-off 
at low effective temperatures.}
\end{figure}

Finally, we list  in Table 1 the colour indices  for our 0.169, 0.242,
0.327 and 0.406 \msun He WD models at some selected low \teff
values.   In addition,  for  each  stellar mass  and \teff value,  we
provide  the surface  gravity ($g$),  the age  (in Gyr),  the absolute
visual  magnitude ($M_V$)  and the  bolometric correction  ($BC$). The
latter is calculated according to (see Bergeron, Wesemael \& Beauchamp 1995)

\begin{equation}
BC= 2.5 \log \int_0^{\infty}{H_{\lambda} S_{\lambda}^V d\lambda} - 10 \log
T_{eff} + 15.6165,
\label{corrbol}
\end{equation}

\noindent where  $H_{\lambda}$ and $S_{\lambda}^V$  are the monocromatic
emergent flux  at temperature \teff  and the transmission  function of
the $V$ filter, respectively.

\section{Conclusions} \label{sec:conclusion}

The present study was aimed  at exploring the evolution of 
white dwarf (WD) stars with helium cores (He WD)
on  the  basis of very  detailed  non-gray  model
atmospheres (Rohrmann 2000) to derive accurate boundary conditions for
the  evolving models.  The emphasis  has been  placed on  the advanced
stages  of  the evolution  of  these  objects  where WD  evolution  is
markedly  dependent   on  the   treatment  of  the   atmosphere.   The
evolutionary phases leading  to the formation of cool  He WDs have
been explored  in detail by Althaus  et al. (2000a),  who found that
the  inclusion of  element diffusion  in evolutionary calculations for
He WDs leads to hydrogen  envelopes thin enough for stable nuclear
burning to  play a minor  role at late  stages of evolution.   Here we
improved the  above-mentioned calculations by  following the evolution
of He WD  models in a self-consistent way  with the predictions of
non-gray model  atmospheres. Another motivation for  the present work
has been to construct a theoretical luminosity function for He WDs
and  to   provide  colour  indices   and  magnitudes  for   these  WDs
self-consistently with stellar evolution and element diffusion.

In particular, we have considered He WD models with stellar masses
of  0.406, 0.360,  0.327, 0.292,  0.242,  0.196 and  0.169 \msun,  the
evolution  of  which has  been  followed from  the  end  of mass  loss
episodes  during  the  pre-WD  evolution  down  to  very  low  surface
luminosities.

We find that  when \teff decreases below 4000K,  the emergent spectrum
of He WDs becomes bluer  and not redder, as reported by other
investigators. We also analyse the evolution of our He WD models in
the colour-magnitude diagrams and we find that He WDs with masses ranging
from $\sim$  0.18 to 0.3 M$_{\odot}$  can reach the  turn-off in their
colours and become blue again  within cooling times much less than 15
Gyr  and then remain  brighter than  $M_V \approx$  16.5.  This  is an
interesting result  because it raises the possibility  that many low-mass
He WDs  could have had time  enough to evolve  beyond the turn-off
point and  present blue  colours. Cooling times  this short  have also
been  derived by  Althaus et  al (2000a,b)  on the  basis  of detailed
evolutionary models that consider the pre-WD evolution and the effects
of element  diffusion.  As  shown by these  authors, these  new models
solve the discrepancy between the predictions of standard evolutionary
calculations (which do not  consider element diffusion, such as Driebe
et al. 1998)  for He WDs and the age  of some millisecond pulsar
companions. The  detection  of blue,  low-mass He WDs  at very low  
\teff would place the  theoretical  predictions of  our  models  on  a  
firm observational basis.

Complete tables containing the results of the present calculations are
available  at   http://www.fcaglp.unlp.edu.ar/$\sim$althaus/  or  upon
request to the authors at their e-mail addresses.

We thank our anonymous referee whose comments and suggestions greatly
improved the original version of this paper.

\bsp

\label{lastpage}

\end{document}